\documentclass[prl,twocolumn,showpacs]{revtex4}
\usepackage{graphicx}
\begin{document}

\title{Universal and wide shear zones in granular bulk flow}
\author{Denis Fenistein} \affiliation{Kamerlingh Onnes Laboratory,
Leiden University, PO box 9504, 2300 RA Leiden, Netherlands}
\author{Jan Willem van de Meent}\affiliation{Kamerlingh
Onnes Laboratory, Leiden University, PO box 9504, 2300 RA Leiden,
Netherlands}
\author{Martin van Hecke}\affiliation{Kamerlingh Onnes Laboratory, Leiden
University, PO box 9504, 2300 RA Leiden, Netherlands}
\date{\today}
\begin{abstract}

We present experiments on slow granular flows in a modified
(split-bottomed) Couette geometry in which wide and tunable shear
zones are created away from the sidewalls. For increasing layer
heights, the zones grow wider (apparently without bound) and evolve
towards the inner cylinder according to a simple, particle-independent
scaling law. After rescaling, the velocity profiles across the zones
fall onto a universal master curve given by an error function. We
study the shear zones also inside the material as function of both
their local height and the total layer height.

\end{abstract}

\vspace{-0.1mm} 

\maketitle

Slowly sheared granular matter does not flow homogeneously like a
liquid. Instead, granulates form rigid, solid-like regions
separated by narrow shear bands where the material yields and
flows
\cite{duran,nedderman,jaegerrmp,jaegerscience,hartley,mueth}.
Shear localization is ubiquitous in granular flow --- think of
geological faults and soil fractures, avalanches, pipe flows and
silo discharges
\cite{scott,oda,bridgewater,mulhaus,daerr,komatsu,nedderman,neddermanlaohakul,pouliquen}.

Despite their crucial importance, granular shear flows are still
poorly understood, in part because shear localization itself remains
enigmatic \cite{duran,nedderman,jaegerrmp,jaegerscience}.  On the one
hand, shear bands have a typical thickness of five to ten grain
diameters and such steep gradients are difficult to capture by
continuum theories
\cite{duran,nedderman,jaegerrmp,jaegerscience,hartley,mueth,neddermanlaohakul,pouliquen,scott,oda,bridgewater,mulhaus,daerr,komatsu}.
On the other hand, the experimental handles for probing shear
localization are limited. For example, studies in Couette cells always
show the formation of a narrow shear band near the inner cylinder,
irrespective of dimensionality, driving rate or details of the
geometry \cite{hartley,mueth,howell,losert1,losert2,latzel}: Shear
banding is very robust.

\begin{figure}[t]
\includegraphics[width=8cm]{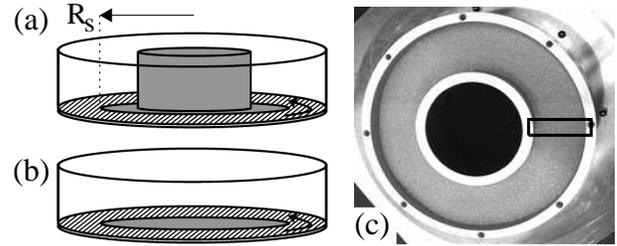}
\caption{(a) Schematic side-view of our split-bottomed Couette cell,
showing the stationary bottom disk and inner cylinder (both grey), the
rotating bottom ring (striped) and outer cylinder. A thin layer of
felt between the rings assures smooth rotation. The inner and outer
cylinder radii are fixed at $65$ mm and 105 mm respectively, while
$R_s$ can be varied. (b) Setup in disk geometry. (c) Top view of the
filled set up in the Couette geometry, where the rectangle indicates the area
recorded by the video camera. }\label{fig1}
\end{figure}

In this Letter, we introduce a general experimental protocol that can
yield very wide shear zones away from the side-walls.  We modify a
Couette cell by splitting its bottom at radius $R_s$.  The resulting
concentric rings are attached to the stationary inner and rotating
outer cylinder, respectively, and the cell is filled with grains up to
height $H$ (Fig.~\ref{fig1}). When driving the system, a shear zone is
found to propagate from the slip position $R_s$ towards the surface
where we measure the average grain velocities. Note that our strategy
differs from previous works, which were carried out for large filling
heights and smooth bottoms so as to minimize the effect of the bottom
boundary \cite{mueth,losert1,losert2}. Here we take advantage of
gravity and ''drive the system from the bottom''.

Our main finding are: {\em{(i)}} For large $H$, a regime of
wall-localized shear band near the inner cylinder is recovered
\cite{mueth,losert1,losert2}, but for intermediate $H$, we observe
shear zones of tunable width away from the boundaries. This paper
focuses on describing these ``bulk'' shear zones.  {\em{(ii)}} The
angular velocity profiles $\omega(r)$ of bulk shear zones fall onto a
universal master curve which is best fitted by an error
function. These profiles are therefore fully characterized by two
parameters only: their center position $R_c$ and width $W$. A concise
presentation of these results has appeared in \cite{fenistein}.
{\em{(iii)}} $R_c$ and $W$ depend on $H$, $R_s$ and particle
properties in specific manners.  The center of the shear zone $R_c$
evolves to the inner cylinder with increasing $H$ in a particle
independent manner. The shear zone width $W$ grows continuously with
$H$ and depends on particle size and shape, but not on the slip radius
$R_s$.  {\em{(iv)}} For a given height inside the material, the width
and position of the shear zones depend on the height of the free
surface $H$.

{\em Setup -- } A sketch of our split-bottomed Couette cell is shown
in Fig.~\ref{fig1}a.  When the inner cylinder of the ``Couette''
geometry is removed we obtain the ''disk'' geometry
(Fig.~\ref{fig1}b). Different sets of bottom rings allow us to vary
$R_s$ from 45 mm to 95 mm. Grains, similar to those used in the bulk,
are glued to the side walls and bottom rings to obtain rough
boundaries. We studied spherical glass beads of size distributions
$0.25-0.42$ mm (I), $0.56-0.8$ mm (II), $1-1.2$ mm (III) and $2 -2.4$
mm (IV), and irregularly shaped plastic flakes ($1.0-1.6$ mm) (V),
aluminum oxide beads ($1.5 -2$ mm) (VI) and coarse sand ($1.2 -2.4$
mm) (VII). After filling the cell, an adjustable blade flattens the
surface at the desired height. The outer cylinder and its co-moving
ring are then rotated. A Pulnix TM-6710 8-bit CCD camera records
2000-frame movies of the resulting flow at the top surface at a rate
of 120 frame/s with pixel resolution 100 $\mu$m.

\begin{figure}[t]
\includegraphics[width=8cm]{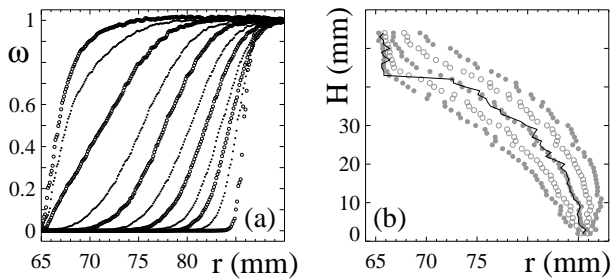}
\caption{ Main features of the normalized angular surface velocity
in the Couette geometry for 0.3 mm glass beads (mixture I) and $R_s =
85$ mm. (a) $\omega(r)$ for a range of equidistant heights
$h=3,8,\dots,53$ mm (right to left). (b) Contour plots of
$\omega(r)$, where the symbols correspond to, from left to right,
$\omega=0.1,0.25,0.5,0.75$ and $0.9$. The curve indicates the strain
rate maximum and shows the rapid qualitative change of the profiles
when the inner cylinder is approached.}  \label{fig2}
\end{figure}

The flow rapidly ($\sim 1$ s) reaches a stationary state where it is
purely in the azimuthal direction, so that the surface velocities is a
function of the radial coordinate only
\cite{hartley,mueth,howell,losert1,losert2,fenistein}. We checked that
these velocities are proportional to the driving rate $\Omega$
\cite{mueth,losert1,latzel,fenistein} for $0.16<\Omega<1.5$ rad/s, and
subsequently fix $\Omega$ at 0.16 rad/s. We thus focus on the velocity
profile $\omega(r)$, the dimensionless ratio of the average angular
velocity and $\Omega$. We measure $\omega(r)$ with high radial
resolution by particle image velocimetry, i.e., by determining the
averaged angular correlation function as function of $r$ of two
temporally separated frames. Unless noted otherwise, the time
separation between frames is around $0.3s$.

{\em Basic phenomenology -- } Figure~\ref{fig2} illustrates the
main features of these velocity profiles. For shallow layers, a
narrow shear zone develops above the split at $R_s$. When $H$ is
increased, the shear zone broadens continuously and without any
apparent bound. The broadest observed zones exceed 50 grain
diameters in width. Additionally, with increasing $H$, the shear
zone shifts away from $R_s$ towards the center of the shear cell
\cite{1overr}. Indeed, for sufficiently large $H$, the shear zone
reaches the inner wall, where it approaches the asymptotic regime
of wall-localized shear bands reported earlier
\cite{mueth,losert1,losert2}. Before this wall localization
occurs, however, there is a substantial range of layer heights
where wide and symmetric bulk shear zones can be observed.

\begin{figure}[t]
\includegraphics[width=8cm]{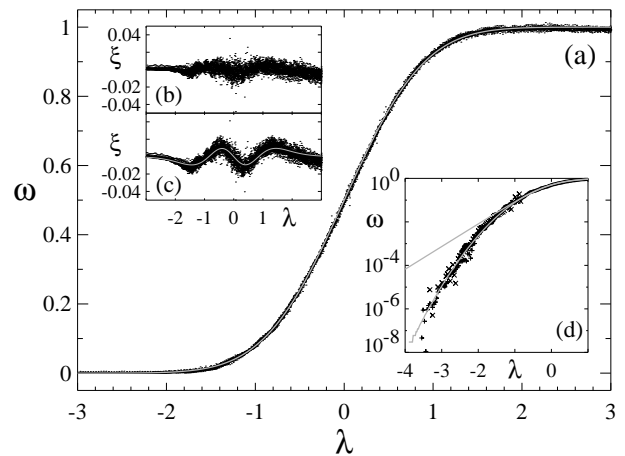}
\caption{(a) Collapse of all 35 bulk profiles obtained for $H=4,5,6,
\dots 38$ mm (mixture I, $R_s=85$ mm, as in Fig.~\ref{fig2}) when
plotted as function of the dimensionless coordinate $\lambda :=
(r-R_c)/W$. The grey curve is an error function.  (b) Differences $\xi$
between bulk profiles and fits to Eq.~(\ref{universal}). (c)
Differences $\xi$ between bulk profiles and fits to $1/2+1/2 \tanh(l
\lambda)$, where $l \approx 1.20274$ compensates for the difference in
scale between error function and hyperbolic tangent. The grey curve
shows $ (\mbox{erf}(\lambda)-\tanh(l\lambda))/2$.  (d) Left tail of
the velocity profile in log-lin scale, compared to the best fit to
either a hyperbolic tangent and error function (grey curves).  Dots,
crosses and pluses are obtained for time-lags 0.3, 3, and 30 s
respectively. }\label{fig3}
\end{figure}

{\em Universal velocity profiles -- } Figure~\ref{fig3} illustrates
our main result: After proper rescaling, all bulk profiles collapse on
a universal curve which is extremely well fitted by an error function:
\begin{equation}\label{universal}
\omega(r)= 1/2+1/2 \mbox{ erf}~((r-R_c)/W)~.
\end{equation}

A residue analysis comparing the fit to Eq.~(\ref{universal}) to an
alternative fit to an hyperbolic tangent shows that the fit to the
error function is always better (Fig.~3b-c). By repeating this
procedure for the other particle mixtures, we establish the general
superiority of Eq.~(\ref{universal}): {\em particle shape does not
influence the functional form of the velocity profiles}. The robust
form of $\omega(r)$ contrasts with the particle dependence found for
wall localized shear bands \cite{mueth}. For these, the vicinity of
the wall causes layering, in particular for monodisperse
mixtures. Apparently such layering effects play no role for our bulk
shear zones. Accurate measurement of the tail of the velocity profile
(Fig.~\ref{fig3}d) further validate Eq.~(\ref{universal}), and rule
out an exponential tail of the velocity profile here. The strain rate
is therefore Gaussian, and the shear zones are completely determined
by their centers $R_c$ and widths $W$.

What limits the universal regime? Apart from wall-localization (see
Fig.~\ref{fig2}), we find that in the disk geometry $\omega(r)$ starts
to deviate from Eq.~(\ref{universal}) when $H$ exceeds $\sim
R_s/2$. The symmetry of the velocity profile, easily detectable by a
simple $\chi^2$ test, is then weakly broken \cite{symnote}. In the
following, we focus on the functional dependencies of $R_c$ and $W$ on
the parameters $R_s$, $H$ and particle type for the universal profiles
given by Eq.~\ref{universal}.

\begin{figure}[t]
\includegraphics[width=8cm]{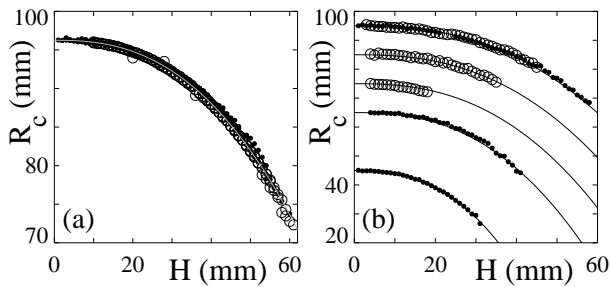}
\caption{Shear zone positions $R_c$ versus height, where $H$ is
restricted to the universal regime detected by a $\chi^2$ test.  (a)
Comparison between $R_c(H)$ for glass beads I-IV (closed symbols), non
spherical particles V-VII (open symbols) and Eq.~(\ref{rseq})
(curve). For all runs, $R_s$ is 95 mm. (b) $R_c(H)$ for mixture III
and $R_s$ = 95, 85 and 75 mm in the Couette geometry (open symbols)
and $R_s$ = 95, 65 and 45 mm in the disk geometry (closed symbols),
compared to Eq.~2 (curves).}\label{fig4}
\end{figure}

{\em Shear zone position-- } Remarkably, the shear zone center
evolution with height, $R_c(H)$, turns out to be independent of the
grain properties (Fig.~\ref{fig4}a). Therefore, the only relevant
length-scales for $R_c$ are the geometric scales $H$ and $R_s$. The
dimensionless displacement of the shear zone, $(R_s-R_c)/R_s$, should thus
be a function of the dimensionless height ($H/R_s$) only. The simple
relation
\begin{equation}\label{rseq}
 (R_s-R_c)/R_s = (H/R_s)^{5/2}~
\end{equation}
fits the data well (Fig.~\ref{fig4}a) \cite{parkeetnote}. To check
Eq.~(\ref{rseq}) we have varied $R_s$ over a substantial range. Only
the presence of the inner cylinder limits the range of $R_s$. We find
no differences between bulk velocity profiles measured with or without
the inner cylinder --- {\em Bulk shear zones are insensitive to the
presence of the side walls}. So, we subsequently switched to the disk
geometry (Fig.~\ref{fig1}b), and obtained an excellent agreement
between $R_c$ and Eq.~(\ref{rseq}) over the range $45 < R_s < 95$ mm (
Fig.~\ref{fig4}b).

\begin{figure}[t]
\includegraphics[width=8cm]{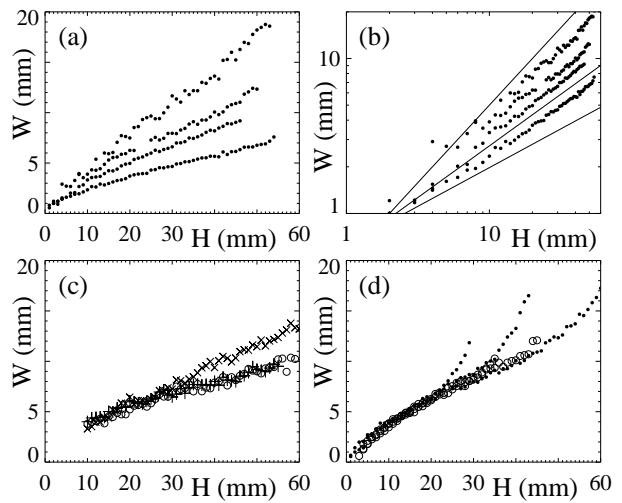}
\caption{ Width of the shear zones versus $H$. (a) Linear and (b)
log-log plots of $W$ for spherical glass beads of increasing sizes for
$R_s\!=\!95$ mm (Mixture I-IV). The width grows with particle
size. The straight lines in (b) have slopes 1/2, 2/3 and 1, showing
that $W$ grows faster than $\sqrt{H}$ but slower than $H$. (c)
Irregular particle shapes diminish $W$ substantially, as can be seen
by comparing the widths for the plastic flakes (V/pluses), aluminum
oxide beads (VI/open circles) and coarse sand (VII/crosses) to those
of the glass beads shown in panel (a). (d) $W$ for Mixture III and
$R_s$=95, 85 and 75 mm (Couette geometry; open symbols) and $R_s$=95,
65 and 45 mm (disk geometry; closed symbols). The strong upward
deviations observed in the disk geometry coincide with the symmetry
breaking of the velocity profiles (see text). Apart from this, the
setup geometry does not affect $W$.  }\label{fig5}
\end{figure}

{\em Shear zone width -- } The width of the shear zones depends on the
particle size and type (Fig.~\ref{fig5}a-c), but not on $R_s$
(Fig.~\ref{fig5}d). First of all, $W$ grows with $H$ and increases for
larger particles (Fig.~\ref{fig5}a-b). The data shown in
Fig.~\ref{fig5}a-b can be made to collapse when plotted as $W/d$ vs
$H/d$ (not shown), where $d$ denotes the grain size.  Grain shape and
type also influences $W(H)$: irregular particles display narrower
zones than spherical ones of similar diameter (Fig.~\ref{fig5}c).
Finally, for the universal velocity profiles, $W$ is independent of
$R_s$ (Fig.~\ref{fig5}d). We therefore conclude that the relevant
length-scale for $W$ is given by the grain properties.

The evolution of the velocity profiles from a step function at the
bottom to an error function at the surface, is reminiscent of a
diffusive process along the vertical axis. However, $W$ grows faster
than $\sqrt{H}$ as diffusion would suggest, but slower than
$H$. Intriguingly, we obtain the best fit for $W\propto H^{2/3}$ over
the limited range where we have reliable data. We cannot, however,
rule out other functional dependencies such as a crossover from
square-root to linear behavior.

\begin{figure}[t]
\includegraphics[width=8cm]{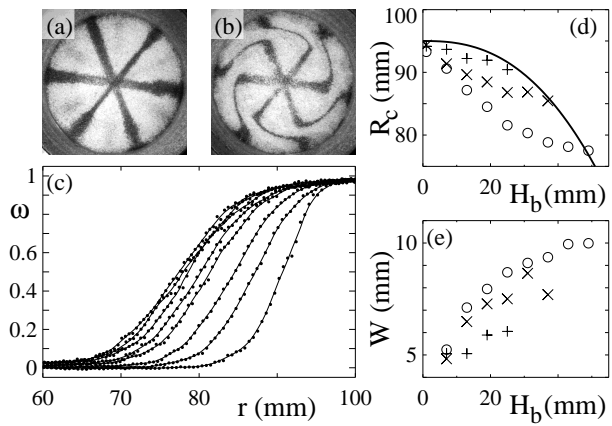}
\caption{Measurements of $\omega(r)$ inside the material for the disk
geometry and $R_s=95$ mm. (a) Initial pattern at $H_b=31$ mm. (b) The
same pattern after adding material up to $H=49$ mm, rotating and
removing the upper layers. (c) Velocity profiles for $H=49$ mm, and
$H_b=7,13,19,\dots,49$ mm (from right to left). (d) $R_c(H_b)$ for
$H=25$ mm (pluses), $H=37$ mm (crosses) and $H=49$ mm (circles) compared
to $R_c(H)$ given by Eq.~(\ref{rseq}) (curve). (e) Corresponding
$W(H_b)$.  }\label{fig6}
\end{figure}

{\em Below the surface ---} So far we have only discussed observations
of the surface flow. To get some insight into the 3D bulk flow
structure, we put patterns of lines of colored tracer particles at
given $H_b$ (Fig.~\ref{fig6}a). More material is carefully added so
that the line-pattern is buried under a given amount of grains ($H >
H_b$). We then rotate the system for a short period ($sim 8$ s), and
recover the deformed line-pattern by carefully removing the upper
layers of grains (Fig.~\ref{fig6}b). Comparing the snapshots of the
deformed pattern to the initial one allows for the determination of
the velocity profiles {\em in the 3D bulk of the material}
(Fig.~\ref{fig6}c).  We have checked that transient effects are
limited, that the measurements reproduce well and that there is no
significant motion in the vertical direction.

The position of the shear zones in the 3D bulk are presented in
Fig.~\ref{fig6}d. Clearly, the evolution of $R_c$ with $H_b$ inside
the material depends on the total amount of matter, as given by $H$:
the more material is added, the more the shear zone shifts towards the
center. This observation is confirmed by recent theory \cite{tamas},
numerics \cite{stef} and MRI measurements \cite{paul}.
The widths of the 3D shear zones are more difficult to measure
accurately, but a clear trend can be identified: shear zones become
wider when more matter is added on top (Fig.~\ref{fig6}e).

{\em Outlook ---} In this work we have presented a simple experimental
protocol in which wide and tunable shear zones can be generated in a
variety of granular materials. Perhaps the biggest surprise is the
robust and remarkably simple form of the velocity profiles --- for
granular systems, universality is rare. Our measurements of the tail
of the profile indicate that even far away from the shear zone, the
grains are not entirely at rest. These findings indicate that these
features are amenable to a continuum description of granular matter
\cite{duran,nedderman,jaegerrmp,jaegerscience}, in particular for very
wide shear zones.

It is also noteworthy that the functions for the shear zone location
$R_c(H)$ and width $W(H)$ depend on entirely distinct sets of
parameters: the relevant length-scales for $R_c$ and $W$ appear to be
well separated. The particle-dependence of the width provides a
characteristic length-scale which may bridge microscopic and
coarse-grained descriptions. The development of theories of granular
flows can further be guided by the universal relation for the shear
zone position, and should incorporate the strong influence of the
boundary: Avoiding the proximity of the side
wall can turn the shear zones from narrow to wide, and from particle
dependent to universal.

The broad shear zones occuring in our geometry allow for further
experiments that are more difficult to realize in narrowly localized
shear bands. Important issues can thus be tackled, such as the
velocity fluctuations and particle diffusion for various strain rates
and locations within the shear zone. One can also probe whether
``local clusters'' of particles, possibly similar to those found in
more rapid flows, would occur for these denser flows
\cite{cluster}. Finally, over the range studied, the velocity profiles
are rate independent, but what happens for much larger and smaller
rotation rates is an open question.  The simple experimental protocol
that we provide for creating generic (i.e., away from sidewalls) shear
zones can be the starting point for many crucial experiments, thus
addressing the basic question: ``How does sand flow?''.

{\em Acknowledgments} We like to thank Floris Braakman for assistance
with the bulk measurements. Financial support by the ``Nederlandse
Organisatie voor Wetenschappelijk Onderzoek (NWO)'' and by ``Stichting
Fundamenteel Onderzoek der Materie (FOM)'' is gratefully acknowledged.


\begin{thebibliography}{99}
\bibitem{symnote} Note that an asymmetric correction to
Eq.~(\ref{universal}) also occurs when one assumes that the shear rate
$\dot{\gamma}(r)$ is strictly symmetric (Gaussian), and calculates
$\omega(r)$ as $\int{r\dot{\gamma}(r)dr}$. This correction to
Eq.~(\ref{universal}) is, however, neglegibly small for our values of
$H$ and $R_s$, so we have ignored it here. More importantly, such
correction grows continuously with $H$, rather than suddenly appearing
at $H \approx R_s/2$.
\bibitem{parkeetnote} We found some indications that strongly
elliptical grains (birdseed/not presented here) deviate from Eq.~2.
\end{thebibliography}
\end{document}